\documentclass[aps,reprint,groupedaddress,longbibliography]{revtex4-1}

\usepackage{graphicx}% Include figure files
\usepackage{dcolumn}% Align table columns on decimal point
\usepackage{bm}% bold math
\usepackage{color}

\usepackage[utf8]{inputenc}
\usepackage[T1]{fontenc}
\usepackage{mathptmx}

\begin{document}

\title[Capillary sorting of fiber suspensions]{Capillary sorting of fiber suspensions by dip coating}

\author{Justin Maddox,$^1$ and Alban Sauret$^{1,2}$}
\affiliation{$^1$ Department of Mechanical Engineering, UC, Santa Barbara, CA 93106, USA}
\affiliation{$^2$ Department of Mechanical Engineering, University of Maryland, College Park, MD 20742, USA}
\email{asauret@ucsb.edu}

\date{\today}

\begin{abstract}
Sorting elongated anisotropic particles, such as fibers, dispersed in suspensions poses significant challenges as they present two characteristic dimensions: length and diameter. Fibers in suspension usually align with the flow, leading to diameter-based filtration when passing through a sieve. Modifying the flow conditions by introducing more mixing so that fibers are arbitrarily oriented can lead to sorting by diameter and length simultaneously, resulting in a lower filtration quality. In this paper, we demonstrate that capillary filtration by dip coating can be utilized to selectively sort fibers by length or by diameter in a controlled manner. Using the withdrawal of a flat substrate from a fiber suspension, we demonstrate that fibers are primarily sorted by their diameters. When considering cylindrical substrates, fibers can be sorted by length under appropriate conditions due to the orientation adopted by the fibers during their entrainment. We report guidelines for designing this filtration process and obtaining good sorting efficiency.
\end{abstract}

\maketitle

\noindent \textbf{Introduction.} Filtration of particulate suspensions is critical in many industries as a method of purifying liquids, controlling particle sizes, and separating groups of particles \cite{haw2004jamming,merkus2009sieves,crittenden2012mwh,dressaire2017clogging}. Some of the most commonly used methods are sieving, where the fluid passes through a filter of certain pore size \cite{sauret2014clogging}; centrifugal filtration, where centrifugal force is used to filter particles based on density \cite{gimbert2005comparison}; and cross-flow filtration, where flow is tangential to a membrane and a transmembrane pressure differential is responsible for pulling smaller particles out of the flow \cite{henry2018cross}. These standard filtration methods can struggle with anisotropic particles as they present two lengthscales that can be defined, for instance, through a minimum and a maximum Feret diameter \cite{endoh1984sieving,walton1948feret,duchene2020clogging}.

A potential solution to filter anisotropic particles, specifically fibers, is capillary filtration using a dip coating process. The principle is that a substrate is submersed within a suspension bath, and based on the withdrawal velocity of the substrate and the properties of the interstitial fluid, only particles below a certain threshold size will be entrained in the liquid film coating the substrate after removal \cite{Gans2019,palma2019dip,shovon2022micro,copeland2023dip}. Previous works have demonstrated the efficacy of sorting spherical particles using dip coating with a planar substrate \cite{sauret_2019,dincau2019capillary} and cylindrical substrates \cite{dincau2020entrainment,khalil2022sorting}. These sorting methods rely on the importance of the size of the particle compared to the size of the capillary film. Such interplay between liquid/air interfaces and solid particles have also been considered for liquid thread \cite{furbank2004experimental,bonnoit2012accelerated,chateau2018pinch,thievenaz2021droplet,thievenaz2022onset} and films \cite{raux2020spreading,jeong2022particulate}.

To provide some physical context, after a substrate has been immersed into a liquid bath, upon withdrawal, it becomes coated with a thin liquid film whose thickness, $h$, depends on the withdrawal velocity $U$, the properties of the liquid (surface tension $\gamma$, density $\rho$, dynamic viscosity $\mu$), and the surface properties (geometry, roughness, etc. \cite{krechetnikov2005experimental,rio2017withdrawing}). For a planar substrate and a Newtonian fluid, the Landau-Levich-Dejarguin law predicts:
\begin{equation}
h=0.94\,\ell_{\rm c}\,{\rm Ca}^{2/3}, \label{eq:LLD}
\end{equation}
 where $\ell_{c} = \sqrt{(\gamma/(\rho g)}$ is the capillary length and ${\rm Ca} = \mu\, U / \gamma$ the capillary number \cite{landau1942physicochim,DERJAGUIN1993134}. For cylindrical substrates, the coating thickness involves the Goucher number, ${\rm Go}= R_{s}/\ell_{c}$ \cite{white1965static,white1966theory,Quere1999,rio2017withdrawing} (where $R_{s}$ is the radius of the substrate) and can be estimated through \cite{dincau2020entrainment}:
\begin{equation}
\frac{h}{R_{\rm s}} = \frac{1.34 \,{\rm Ca}^{2/3}}{1+2.53\,{\rm Go}^{1.85}/[1+1.79\,{\rm Go}^{0.85}]} \label{eq:Goucher}
\end{equation}

The filtering of spherical non-Brownian particles has been shown to depend on $h^{*}$ \cite{colosqui2013hydrodynamically,sauret_2019}, which corresponds to the thickness of the liquid at the point where the static meniscus ends and the dynamic meniscus begins, also known as the stagnation point \cite{Mayer2012}. This point also corresponds to the location that separates the flow into two regions: a region that continues into the coating film and a region where the fluid recirculates into the liquid bath. The particles that make it past the stagnation point and that are of size comparable to $h^{*}$ see a sharp increase in frictional forces, which allows for the overcoming of capillary forces \cite{jeong2020deposition}. In the limit of small capillary numbers, the thickness at the stagnation point and the film thickness are related through $h^{*} \approx 3h$ \cite{dincau2020entrainment}. For spherical particles, the sorting by diameter sorting occurs when $h^{*} \lesssim d/2$ \cite{sauret_2019}. From a qualitative point of view, the mechanism associated with the capillary sorting of particles is similar to a dynamic "filter" whose size is governed by the film thickness $h$ (and thus by the thickness at the stagnation point $h^*$). As the substrate is withdrawn from the suspension bath, the liquid film creates a "channel" that particles must enter to become entrained, bounded on one side by the substrate and on the other side by the air-liquid interface \cite{dincau2019capillary}.

In the case of fiber suspensions, the situation becomes more complex as, in addition to the diameter $d$, a new lengthscale is introduced: the length $L$. A recent study suggested that for dip coating with cylindrical substrates, fibers of length larger than the diameter of the substrate were entrained only at larger withdrawal velocities than predicted based on their diameter \cite{jeong2023deposition}. This observation suggests that using different kinds of substrates (flat or cylindrical) could potentially control whether sorting fiber suspensions by capillary filtration occurs through the diameter $d$ or the length of the fibers $L$ and could thus provide a unique sorting platform. In this paper, we examine if such a filtration method is possible, and we report the conditions under which this could be realized. We first present our experimental methods before considering the filtration by diameter with a flat substrate. In a second time, using a cylindrical substrate, we report the condition and the efficiency of length-based filtration. Finally, we discussed the limits of this method and how to optimize the filtration process.

\bigskip

\begin{figure}
    \centering
    \includegraphics[width=0.9\linewidth]{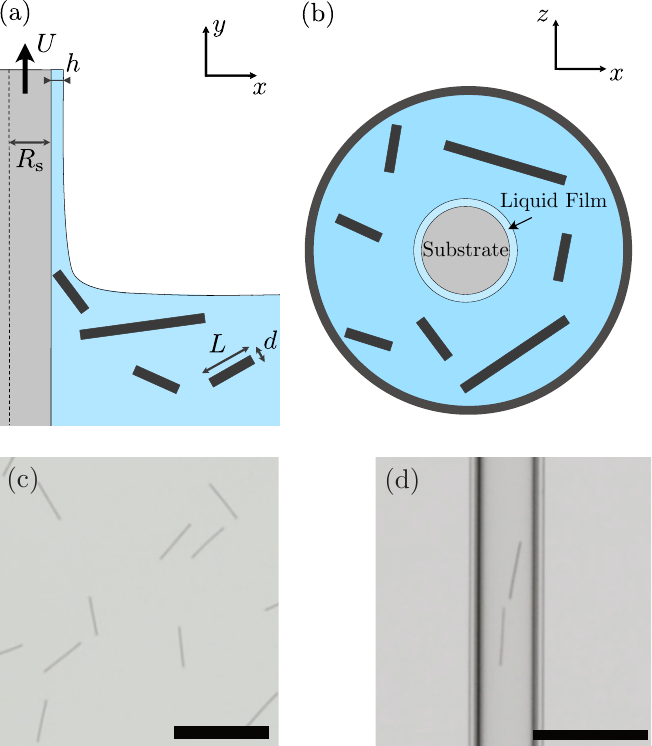}
    \caption{(a) Side and (b) top view schematics of the dip coating process with a suspension of fibers of diameter $d$ and two different lengths $L_1$ and $L_2$. Note that the size of the fibers, the substrate, and the meniscus are not to scale. (c)-(d) Examples of entrainment of fibers on (c) a flat substrate where no preferential orientation is observed and (d) on a rod where the fibers entrained are of length comparable to the substrate diameter and thus mainly align with the axis of the rod.}
    \label{fig:figure_1_setup}
\end{figure}

\noindent \textbf{Methods.} Schematics of the experimental setup are shown in Figs. \ref{fig:figure_1_setup}(a)-(b). The suspensions are made of non-Brownian nylon fibers of diameter $200\,\mu{\rm m} \leq d \leq 400\,\mu{\rm m}$ and length $1.5\,{\rm  mm} \leq L \leq 4.5 \,{\rm  mm}$ dispersed in Fluorosilicone (FMS-221, Gelest), a Newtonian fluid of viscosity $\mu_{\rm f} =  109\, {\rm mPa.s}$, surface tension $\gamma = 21 \, \mathrm{mN.m^{-1}}$ and density $\rho_{\rm f} = 1160 \, \mathrm{kg.m^{-3}}$. The fibers used are cut to the desired length by mechanical chopping (WC302 Automatic Wire Cutter, The Eraser Company), resulting in a well-controlled average length $L$ and a standard deviation on the length of $0.25\,{\rm  mm}$. Due to the mechanical chopping, the edges of the fibers are not perfectly cylindrical, but, for instance, a little bit flattened irregularly. Nevertheless, we did not observe any effects of the edge geometry on the sorting process. In our experiments, no deformation of the fibers was observed for any aspect ratio considered. The density of the liquid is close enough to the density of the fibers ($\rho \simeq 1150  \,\mathrm{kg.m^{-3}}$) so that buoyancy effects can be neglected over the timescale of the experiments. The volume fraction of fibers used, $\phi = V_{\rm fib}/V_{\rm tot} = 0.5\%$, ensures that we remain in the dilute regime \cite{butler2018microstructural}. The suspensions are initially mixed overnight using magnetic stirrer bars and mixed for 5 minutes between trials. Using optical absorbance, we ensured that our experimental protocol led to fibers initially homogeneously dispersed in the suspension bath.

We perform the experiments with two substrate geometries: a planar substrate that consists of a glass plate (7mm thick, 50mm wide, 70 mm length; see figure \ref{fig:figure_1_setup}(c)) and a cylindrical substrate setup consisting of an array of three rods (each of radius $R_{\rm s}=1.2\,{\rm mm}$; see figure \ref{fig:figure_1_setup}(d)), withdrawn simultaneously from the suspension bath. The substrates are withdrawn from the suspension bath using a linear translation stage (Thorlabs NRT150 linear stepper motor) at a velocity ranging from $2.5 \,{\rm mm.s^{-1}}$ to $20 \,{\rm mm.s^{-1}}$ over a distance of $30\,{\rm mm}$. In both cases, between trials, the substrates were cleaned using isopropyl alcohol, rinsed with Deionized (DI) water, and dried using compressed air. 

After each trial, the number of particles entrained was measured visually by examining the substrate over the entire length that had been withdrawn from the liquid. We should emphasize that over the time scale of the experiment and the measurement of the fiber entrained, we did not observe any significant sliding down of the fibers along the substrate. We performed experiments with a single population of fibers to determine the threshold capillary number corresponding to each fiber geometry and to study how the number of particles entrained depends on the withdrawal velocity. For the experiments with mixed suspensions, the liquid bath contains two different fiber geometries, each at an equal volume fraction. The results of these tests were used to determine the efficiency of sorting two different fiber lengths or diameters. In the following, each data point reported in the different figures consists of 60 dip coating trials, \textit{i.e.}, withdrawal of the substrate, and counting the number of fibers entrained.

We introduce different parameters to describe our results. The threshold Capillary number, ${\rm Ca^{*}}=\mu_{\rm f}\,U^{*}/\gamma$ refers to the threshold value at which a given geometry of fiber begins to be entrained for a specific substrate ($U^{*}$ is the threshold capillary number). We consider here the capillary number based on the viscosity of the interstitial fluid $\mu_{\rm f}$ since the volume fractions $\phi$ used in this study are very small, so the change in viscosity can be neglected \cite{butler2018microstructural,bounoua2019shear}. The effects of the substrate size compared to the fiber lengths are captured through a normalized fiber length, $L^{*}=L/2R_{\rm s}$, corresponding to the ratio of fiber length to the substrate diameter (for a flat plate, $L^{*} = 0$). We also define the normalized sorting efficiency as the difference between the mass of 'small' fibers entrained, $m_{\rm S}$, and the mass of 'large' fibers entrained, $m_{\rm L}$ (where 'small' and 'large' can refer to the diameter or to the length depending on the situation) rescaled by the total mass of all fibers entrained:
\begin{equation}
\eta = \frac{m_{\rm S}-m_{\rm L}}{m_{\rm S}+m_{\rm L}}
\end{equation}
Therefore, $\eta=1$ refers to perfect sorting, and $\eta=0$ to no sorting at all. Note that here, the volume fraction of both fiber populations is the same. Finally, we define $\phi_{\rm F}/\phi_{\rm U}$ defined as the average volume fraction of particles entrained in the liquid film on the substrate post-withdrawal rescaled by the unfiltered volume fraction of particles in the suspension bath.

\bigskip

\noindent \textbf{Flat substrate: Sorting by Diameter.} We first consider a suspension made of fibers of similar length ($L=1.5\,{\rm mm}$) but two different diameters $d=200\,\mu{\rm m}$ and $d=400\,\mu{\rm m}$. The fibers are mixed at a 1:1 ratio, occupying a volume fraction $\phi=0.5\%$. Because all fibers were cut from a continuous wire of approximately constant diameter, there was no notable distribution of sizes within each diameter group. Dip coating was performed with a planar substrate at various velocities, and we measured the distribution of fibers found on the substrate post-withdrawal. 

\begin{figure*}
    \centering
    \includegraphics[width=\linewidth]{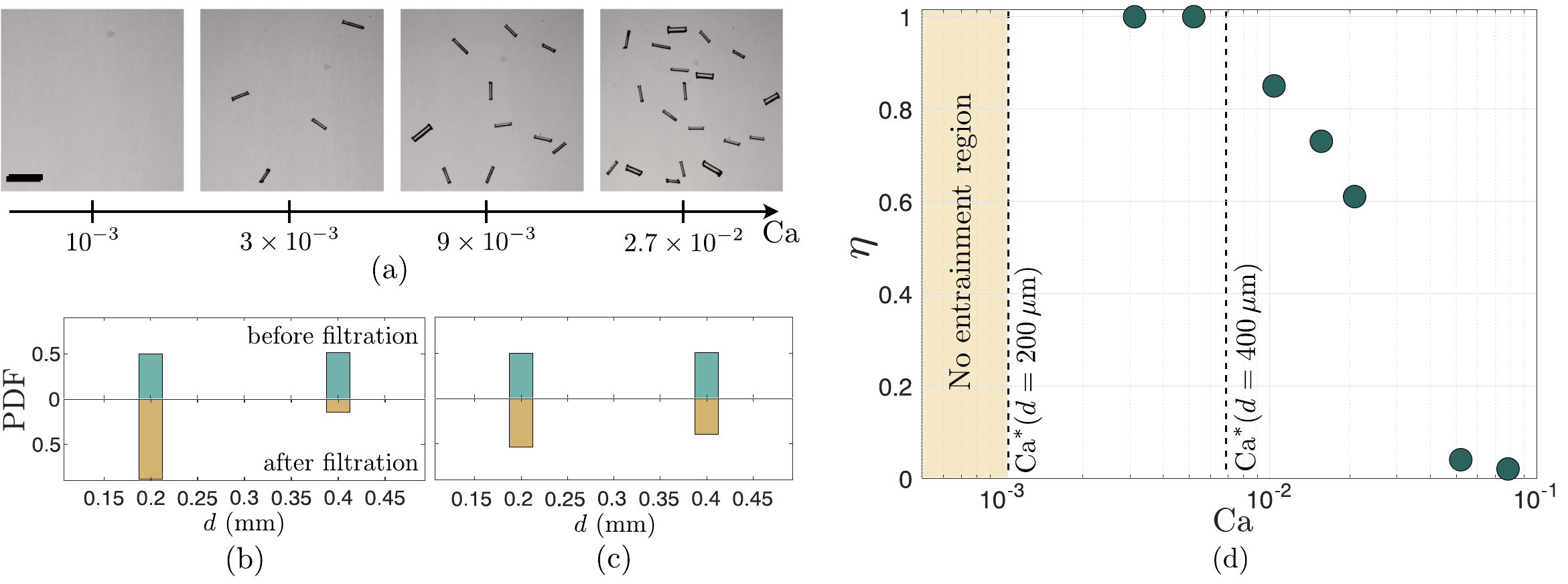}
    \caption{(a) Examples of post-filtration fiber distributions observed on a flat substrate for increasing capillary numbers and an initial suspension made of fibers of length $L=1.5\,{\rm mm}$ and diameter $d=200\,\mu{\rm m}$ and $d=400\,\mu{\rm m}$. From left to right: No entrainment, perfect sorting, high-efficiency sorting, low-efficiency sorting. Scale bar is $3\,{\rm mm}$. (b)-(c) Examples of probability density functions of the fiber diameter before (top PDF) and after (bottom PDF) for (b) a good-efficiency sorting at ${\rm Ca}=10^{-2}$ and (c) low-efficiency sorting at ${\rm Ca}=5 \times 10^{-2}$ for the same suspension than in (a). (d) Evolution of the diameter sorting efficiency $\eta$ as a function of the capillary number with a planar substrate and the same suspension. The shaded area denotes the region with no fiber entrainment. The vertical dashed lines indicate the threshold capillary numbers of each fiber type contained within the suspension. The suspension consists of $L=1.5\,{\rm mm}$ long fibers, with a $0.5\%$ volume fraction of $d=200\,\mu{\rm m}$ diameter fibers and a $0.5\%$ volume fraction of $d=400\,\mu{\rm m}$ diameter fibers.}
    \label{Figure_2}
\end{figure*}

Fig. \ref{Figure_2}(a) shows examples of the results of the filtration process when varying the withdrawal velocity and, thus, the capillary number. At very low values of Ca (${\rm Ca}=10^{-3}$ in the example presented here), no fiber entrainment occurs because each of the two fibers has its own threshold capillary number, $\rm Ca^{*}$, larger than ${\rm Ca}=10^{-3}$, all particles are filtered out. At values of Ca between the individual $\rm Ca^{*}$ values of each fiber, there is theoretically perfect sorting, as seen here for ${\rm Ca}=3 \times 10^{-3}$. At values of Ca above $\rm Ca^{*}$ of both particles  (${\rm Ca}=9 \times 10^{-3}$ in the present example), there are some sorting effects, but its efficiency decreases as Ca increases further. We report in Figs. \ref{Figure_2}(b) and \ref{Figure_2}(c) a more quantitative analysis: the probability density functions of the fiber diameter before (\textit{i.e.}, in the suspension) and after the withdrawal for ${\rm Ca}=10^{-2}$ and ${\rm Ca}=5 \times 10^{-2}$, respectively. As suggested by the observations in Fig. \ref{Figure_2}(a), we observe that the filtration efficiency decreases as the capillary number is increased as fewer fibers with a diameter $d=400\,\mu{\rm m}$ are filtered out [Fig. \ref{Figure_2}(c)].

Fig. \ref{Figure_2}(d) shows that using a planar substrate, it is possible to achieve high efficiencies of diameter sorting between fibers of nearly equal length. Notably, 100\% diameter sorting efficiency was achieved for the trials between ${\rm Ca} \sim 10^{-3}$ and ${\rm Ca} \sim 7 \times 10^{-3}$, which corresponds to capillary numbers between the two thresholds of entrainment for $1.5\,{\rm mm}$ fibers: ${\rm Ca}^*(d=200\,\mu{\rm m}) < {\rm Ca} <{\rm Ca}^*(d=400\,\mu{\rm m})$. At capillary numbers above the threshold value for both fibers, there is a drop in sorting efficiency $\eta$, eventually approaching 0\%, \textit{i.e.}, value associated with complete randomness and no sorting capability, around ${\rm Ca} \simeq 10^{-1}$ for the fibers considered here. In the present case, this value of ${\rm Ca}$ corresponds to a film thickness of order $300\mu{\rm m}$ and a thickness at the stagnation point of around $900\mu{\rm m}$, so that the meniscus does not act as a filter anymore, allowing all the fibers to flow through.

\bigskip

\begin{figure}
    \centering
    \includegraphics[width=\linewidth]{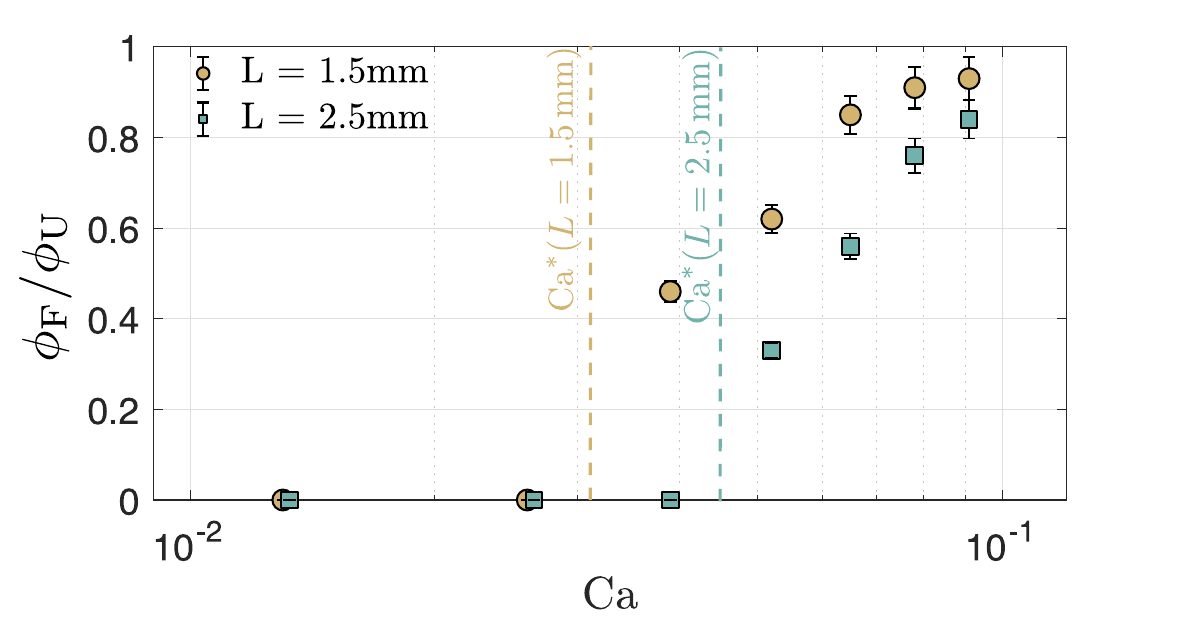}
    \caption{Evolution of the volume fraction of fibers entrained in the liquid film $\phi_{\rm F}$ rescaled by the volume fraction of fibers in the suspension bath, $\phi_{\rm U}=0.5\%$, as a function of the capillary numbers for fibers of diameter $d=200\,\mu{\rm m}$, and length $L=1.5\,{\rm mm}$ (blue circles) and $L=2.5\,{\rm mm}$ (orange squares). The vertical dashed lines indicate the threshold capillary numbers of each fiber type for experiments with only one fiber population in suspension.}
    \label{fig:length_sorting_proof}
\end{figure}

\noindent \textbf{Cylindrical substrate: Sorting by Length.} To explore the possibility of sorting fibers having the same diameters but different lengths, we first performed entrainment experiments using cylindrical substrates of radius $R_{\rm s}=1.2\,{\rm mm}$ and two different fiber lengths, $L=1.5\,{\rm mm}$ and $L=2.5\,{\rm mm}$ separately. The results reported in Fig. \ref{fig:length_sorting_proof} show that, despite having the same diameters, the threshold Capillary numbers are different: $\rm Ca^{*}(L=1.5\,{\rm mm}) \simeq 3 \times 10^{-2}$ and $\rm Ca^{*}(L=2.5\,{\rm mm}) \simeq 5 \times 10^{-2}$. This is due to the fact that $L^{*}=L/(2\,R_{\rm s})$ is different for each population of fibers and the fibers have to reorient to enter the liquid film \cite{jeong2023deposition}. Therefore, theoretically, the ideal range of parameters for sorting these two populations of fibers by length with a cylindrical substrate of radius $R_{\rm s}=1.2\,{\rm mm}$ would be for $3 \times 10^{-2} < {\rm Ca} < 5 \times 10^{-2}$. Fig. \ref{fig:length_sorting_proof} also shows that $\phi_{\rm F}/\phi_{\rm U}$, \textit{i.e.}, the ratio of the volume fraction of fibers filtered rescaled by the volume fraction in the initial suspension increases with ${\rm Ca}$. When $\phi_{\rm F}/\phi_{\rm U} \to 0$, no fibers are entrained on the substrate and when $\phi_{\rm F}/\phi_{\rm U} \to 1$, the meniscus does not filter any fiber, which are then all entrained in the film. Here, we observe that for values of Ca approaching $10^{-1}$, both fiber populations are poorly filtered, each reaching volume fraction ratios of nearly 1, meaning that no separation by length would occur.

\begin{figure*}
    \centering
    \includegraphics[width=\linewidth]{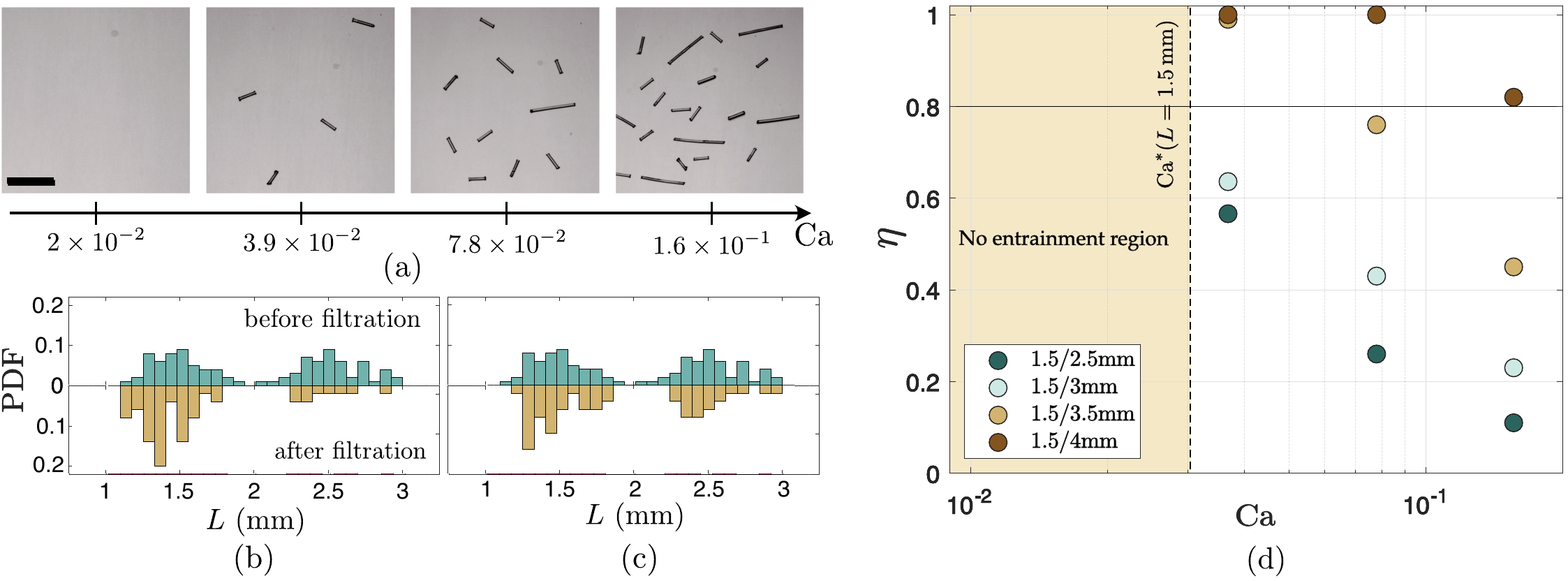}
    \caption{(a) Examples of post-filtration fiber distributions observed on a cylindrical substrate for various capillary numbers. From left to right: No entrainment, perfect sorting, high-efficiency sorting, low-efficiency sorting. Scale bar is $3\,{\rm mm}$. (b)-(c) Examples of probability density functions of the fiber length before (top PDF) and after (bottom PDF) for (b) a good-efficiency sorting at ${\rm Ca} = 3.9 \times 10^{-2}$ and (c) low-efficiency sorting at ${\rm Ca}=7.8 \times  10^{-2}$. In (a)-(c), the initial suspension is made of fibers of $d=200\,\mu{\rm m}$ and average lengths $L=1.5\,{\rm mm}$ and $L=2.5\,{\rm mm}$. (d) Comparison of length sorting efficiency between various fiber lengths, demonstrating a downward trend with increasing capillary numbers. The red area denotes the region of no fiber entrainment. All trials were performed with fibers of $200\,\mu{\rm m}$ diameter and a substrate radius of $R_{\rm s}=1.2\,{\rm mm}$. For all data points, the suspension consisted of 0.5\% of $1.5\,{\rm mm}$ long fibers and 0.5\% of the other length, by volume fraction.}
    \label{fig:figure_4}
\end{figure*}

Following these tests with each fiber geometry, we then considered the possibility to filter a suspension presenting two populations of fibers of $d=200\,\mu{\rm m}$ and average lengths $L=1.5\,{\rm mm}$ and $L=2.5\,{\rm mm}$, respectively. Qualitatively, Fig. \ref{fig:figure_4}(a) shows that filtration is also observed here: (i) no fibers beyond the threshold capillary number of both fiber populations, (ii) a regime where almost only short fibers are entrained, (iii) increasing the capillary number leads to the entrainment of more long fibers as well, until (iv) entraining almost the same composition than the liquid bath at large capillary number (corresponding to $\phi_{\rm F}/\phi_{\rm U} \to 1$ in Fig. \ref{fig:length_sorting_proof}). 

More quantitatively, two examples of the distribution of fiber lengths are shown in Figs. \ref{fig:figure_4}(b) and \ref{fig:figure_4}(c) for ${\rm Ca} = 3.9 \times 10^{-2}$ and ${\rm Ca}=7.8 \times  10^{-2}$, respectively. Before filtration, we observe a bimodal distribution of fiber lengths in both cases (top histograms) peaked around $L=1.5\,{\rm mm}$ and $L=2.5\,{\rm mm}$. The deviation from the mean value comes from the initial cutting of the fibers to obtain the desired lengths. After the withdrawal of the cylindrical substrate, we observe that the distribution skews more towards the $1.5\,{\rm mm}$ fibers for ${\rm Ca} = 3.9 \times 10^{-2}$. We should emphasize that in this case, even with a high sorting efficiency, there are a few long fibers ($L=2.5\,{\rm mm}$) that were entrained on the substrate at a capillary number below its threshold value. This observation is likely due to collective entrainment of particles, as we shall discuss later. At larger capillary numbers, the filtered distribution is almost similar to the initial distribution in the liquid bath [Figs. \ref{fig:figure_4}(c)]. 

To explore a broader range of potential length sorting, the $1.5\,{\rm mm}$ fibers ($L^*=0.63$) were also mixed with $3\,{\rm mm}$, $3.5\,{\rm mm}$, and $4\,{\rm mm}$  fibers, leading to $L^*=1.25$, $1.46$, and $1.67$, respectively. For all the situations, capillary numbers of $3.9 \times 10^{-2}$, $7.8 \times  10^{-2}$, and $1.5 \times 10^{-1}$ were tested. The purpose of these trials is to explore the evolution of the sorting efficiency $\eta$ for different values of $L^*$ and different ratios of fiber length (but same diameter). Fig. \ref{fig:figure_4}(d) showed that, as expected, the trials at ${\rm Ca} = 3.9 \times 10^{-2}$ has a higher efficiency than for larger capillary number, \textit{e.g.}, ${\rm Ca}=7.8 \times 10^{-2}$. Fig. \ref{fig:figure_4}(d) also illustrates that when mixed with $L=4\,{\rm mm}$ fibers, the $1.5\,{\rm mm}$ fibers are perfectly sorted at both ${\rm Ca}=3.9 \times 10^{-2}$ and ${\rm Ca}=7.8 \times 10^{-2}$, indicating that this latter value is still lower than the entrainment threshold of the $4\,{\rm mm}$ fibers. At ${\rm Ca}=1.5 \times 10^{-1}$, none of the mixed suspensions are able to achieve perfect filtration, and the mix of $1.5\,{\rm mm}$  and $2.5\,{\rm mm}$  fibers showed a poor filtration. Considering the different mixes of fibers and a given capillary number, we observe that a suspension of greater length disparity has a higher sorting efficiency than those with two fibers of more similar length. 

Since the threshold capillary number for a given cylindrical substrate increases with the fiber length, in particular when $L^{*}$ becomes larger than unity, the value of $L^{*}$ controls the filtering of a given population of fibers. Low values indicate a decrease in the influence of meniscus curvature around the cylindrical substrate on fiber behavior. The higher the value of $L^{*}$, the tighter the tolerance is for fiber orientation in order to fit within the meniscus and become entrained. Our results suggest that a value of $L^{*} \ge 1$ is reasonable to achieve a good sorting efficiencies. Maximizing $L^{*}$ as much as possible improves the filtering efficiency, but the main drawback to higher $L^{*}$ values is that the threshold withdrawal velocities increase.

\bigskip

\begin{figure}
    \centering
    \includegraphics[width=\linewidth]{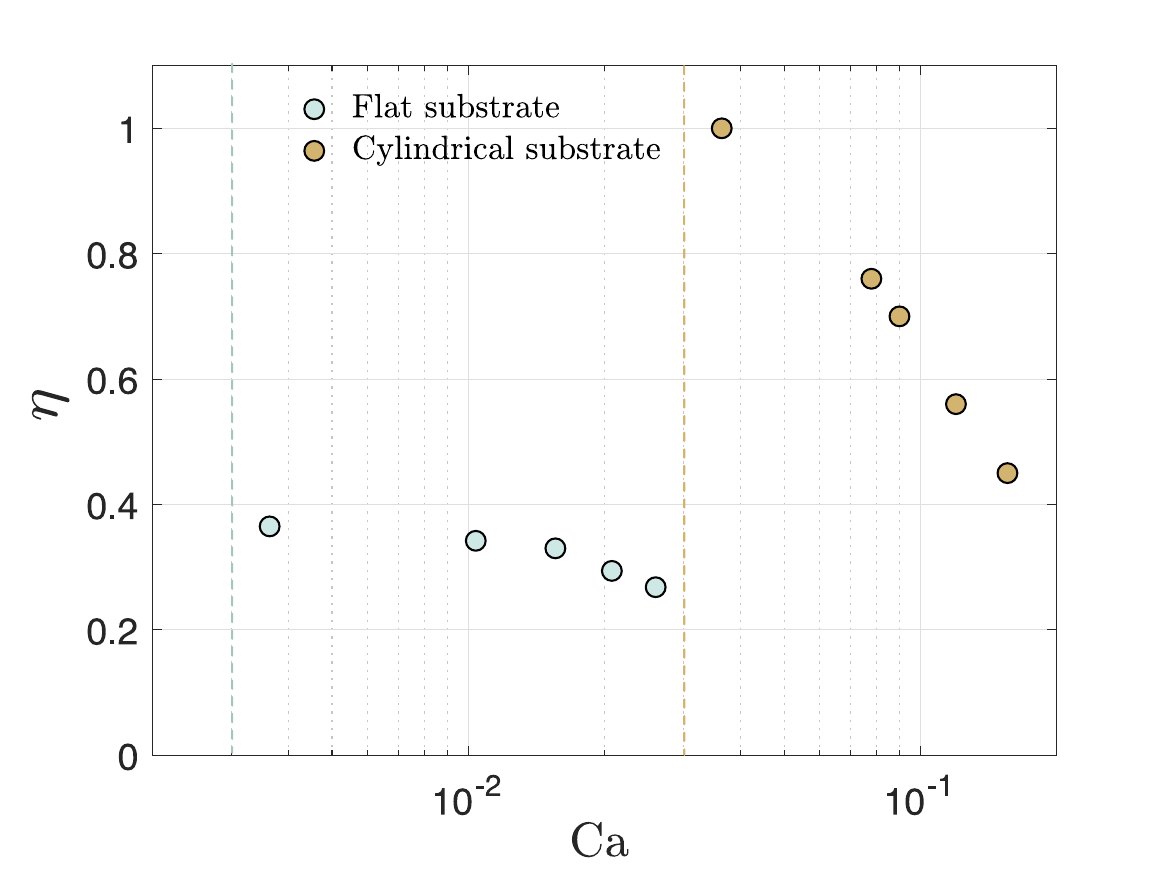}
    \caption{Evolution of the length sorting efficiency $\eta$ for a planar (light green circles) and a cylindrical substrate (brown circles). The suspension consists of 0.25\% of $1.5\,{\rm mm}$ long fibers and 0.25\% of $3.5\,{\rm mm}$ long fibers and of diameters $d=200\,\mu{\rm m}$. The vertical dashed line indicates the capillary threshold for the entrainment of the short fibers for both geometries.}
    \label{fig:figure_7}
\end{figure}

\noindent \textbf{Discussion.} The fibers used here are of length $L \gg h$, so that they will need some amount of reorientation to fit within the meniscus as viscous forces pull them upward or else be subjected to surface forces, which will push them back downward, as described in Ref. \cite{jeong2023deposition}. In addition, a key difference between planar and cylindrical substrates is the curvature in the horizontal direction (around the cylindrical substrates). Past experiments have shown that planar substrates allow for most fibers to have their axis become along the meniscus near the entrainment threshold \cite{jeong2023deposition}, thus allowing the coupling between capillary effects and the frictional force. Conversely, a cylindrical substrate has a curved film, which means that long fibers ($L^* \gtrsim 1$) must reorient such that their main axis becomes parallel to the axis of the substrate to become entrained. This means that for a cylindrical substrate, longer fibers must take a path of higher resistance to entrainment, fully rotating to the preferred orientation, and without an alternative lower resistance path like the planar substrate, longer fibers will be consistently repelled. As a result, the threshold capillary numbers for longer fibers ($L^* \gtrsim 1$) increases with $L^*$. The consequence of this difference is that attempting to sort by length with planar substrates is less efficient, as illustrated in Fig. \ref{fig:figure_7}. However, with cylindrical substrates, a better sorting efficiency can be achieved, as reported in Fig. \ref{fig:figure_7} where when attempting to sort fibers of $200\,\mu{\rm m}$ diameter and length between $1.5\,{\rm mm}$ and $3.5\,{\rm mm}$, a planar substrate (where $L^{*}$ = 0) resulted in a peak efficiency of 36\%, while a $R_{\rm s}=1.2\,{\rm mm}$ radius cylindrical substrate (where $L^{*}$ = 0.6 for the $1.5\,{\rm mm}$ fibers and $L^{*}$ = 1.4 for the $3.5\,{\rm mm}$ fibers) resulted in a peak efficiency of 100\%. We should emphasize that the difference in operating capillary numbers between the two substrates is due to the different evolution of the film thickness with ${\rm Ca}$, as shown in Eqs. (\ref{eq:LLD}) and (\ref{eq:Goucher}) for the flat and cylindrical substrates, respectively.

\begin{figure}
    \centering
    \includegraphics[width=\linewidth]{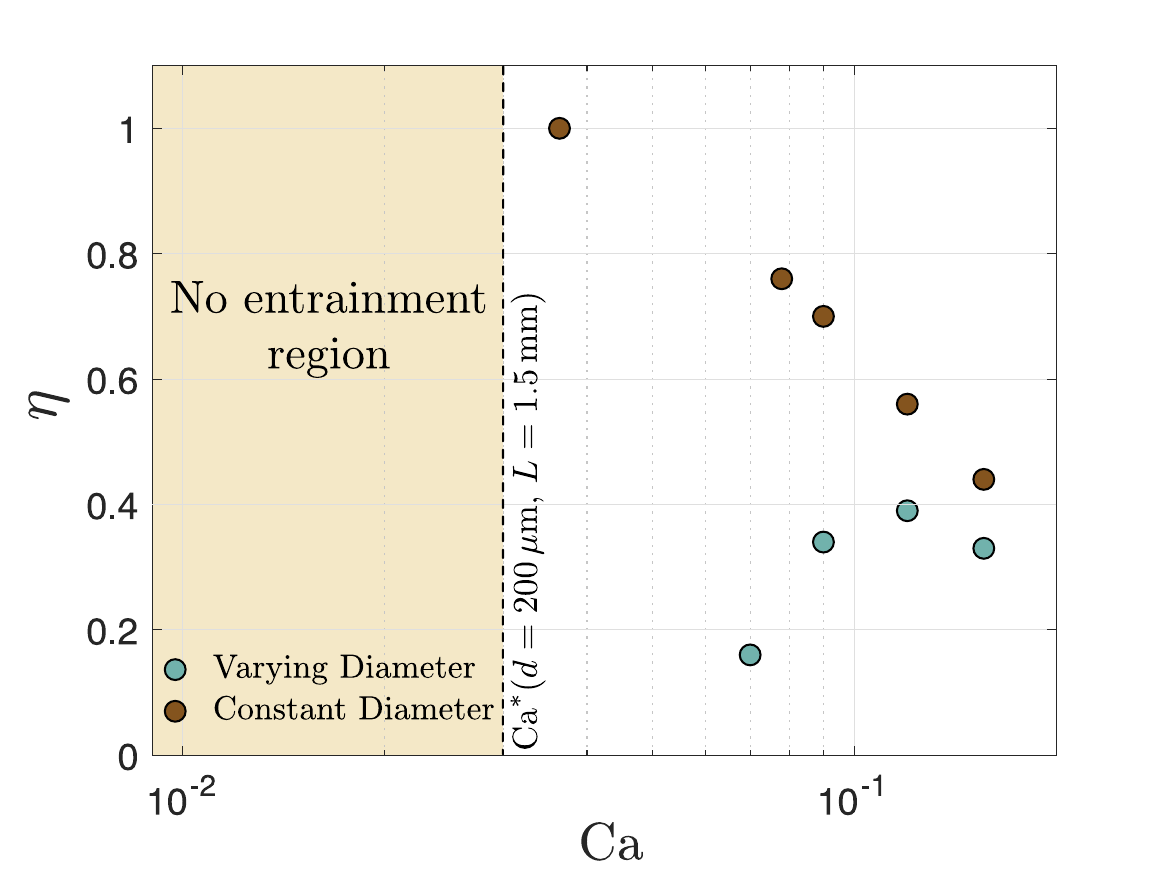}
    \caption{Comparison of the sorting efficiency for various values of ${\rm Ca}$. The varying diameter dataset refers to dip coating experiments with a suspension with volume fractions of 0.25\% of 0.33 mm diameter, $1.5\,{\rm mm}$ long fibers, and 0.25\% of $200\,\mu{\rm m}$ diameter, $3.5\,{\rm mm}$ long fibers. Constant diameter dataset refers to dip coating experiments with a suspension with volume fractions of 0.25\% $200\,\mu{\rm m}$ diameter, $1.5\,{\rm mm}$ long fibers, and 0.25\% $200\,\mu{\rm m}$ diameter, $3.5\,{\rm mm}$ long fibers and serve as the control dataset to observe the relative difference in efficiency when the diameter is varied.}
    \label{fig:figure_5_diameter_length}
\end{figure}

The results presented so far have shown that diameter and length sorting can be performed independently with a very good efficiency, given that the other parameter is held constant. We also performed experiments with fiber populations with both different lengths and diameters: $d=330\,\mu{\rm mm}$ and $L=1.5\,{\rm mm}$ mixed with $d=200\,\mu{\rm m}$ and $L=3.5\,{\rm mm}$. Here, we also consider a case with $d=200\,\mu{\rm mm}$ and $L=1.5\,{\rm mm}$ mixed with $d=200\,\mu{\rm m}$ and $L=3.5\,{\rm mm}$ for comparison. The evolution of the efficiency $\eta$ for both mixed populations is reported in Fig. \ref{fig:figure_5_diameter_length}.

At small values of Ca (${\rm Ca} \lesssim 0.1$ in the example here), a large difference between the efficiencies is observed in Fig. \ref{fig:figure_5_diameter_length}. While the experiments with a constant diameter exhibit a high efficiency, the experiments with two different fiber diameters initially show a very low efficiency, indicating that the diameter sorting effect is roughly of the same order as the length sorting effect. However, for higher values of Ca (${\rm Ca} \gtrsim 0.1$ here), as the ratio of film width to fiber diameter became larger, the diameter sorting effects became negligible for the two diameters considered here, and the two trials reached similar values, showing that the length sorting effect is at play. 

A practical guideline when attempting to sort a suspension by length when different fiber diameters are present would be to first consider sorting using a planar substrate since the diameter sorting effects will usually be of higher order than length sorting effects. In a second time, for the different diameter populations, the sorting could be done with cylindrical substrates as described earlier in this article. If the two-step filtration is not possible, it is desirable to operate at a Capillary number significantly greater than $\rm Ca^{*}$ for the largest diameter in the suspension so that the main filtration effect is by length.

\bigskip

\noindent \textbf{Conclusions.} Using a flat or a cylindrical substrate in a dip coating process, we have shown that it is possible to sort fiber suspensions by length or by diameter. Diameter sorting effects are prevalent when using a planar substrate, typically greatly outweighing length sorting effects. Mechanically, there is no difference in the process of sorting fibers by diameter using a planar substrate or using a cylindrical substrate if $L^* \ll 1$. The diameter sorting process provides a level of reliability that is typically greater than that of length sorting because fibers larger than the threshold size physically cannot pass through the filter; that is to say, it is not susceptible to statistical anomalies. The sorting by diameter is similar to the sorting considered in the past for spherical non-Brownian particles \cite{dincau2019capillary,khalil2022sorting}. Using the different dimensionless quantities introduced in the manuscript (capillary number, Goucher number, $d/L$ and $h/d$ allows to easily adapt the results to different interstitial fluids or fiber sizes.

The theoretical range of optimal efficiency is in the range of capillary numbers between the threshold values of each individual particle suspension. However, it is not always possible to achieve perfect sorting for a mixed suspension, especially if the ratio of the two nominal lengths is close to $1$. Fig. \ref{fig:figure_4} illustrates this point: at Ca = $3.9 \times 10^{-2}$, the sorting efficiency of the mixed $1.5\,{\rm mm}$ and $2.5\,{\rm mm}$ suspension should have theoretically been 100\%, but in practice, it is lower. This observation is similar to observations done in regular filters where some suspensions cannot be sorted perfectly on the first trial because the geometries are too similar. When this is the case, one potential workaround is to sort the same suspension several times, as theoretically, with enough iterations, an effective efficiency of nearly 100\% can be achieved. 

 Even when a fiber suspension has a large enough disparity in length to ideally sort perfectly, volume fraction effects can lead to a decrease in efficiency due to the entrainment of clusters of particles \cite{sauret_2019}. Indeed, when the local volume fraction at the meniscus is too high, collective effects may appear, and the desired length fibers will form clusters that can become tangled with the undesired length and pass through the meniscus as one large aggregate \cite{jeong2022dip}. Reducing the average volume fraction of the entire suspension will help to minimize this effect. Ideally, the volume fraction would be as low as possible, however, very low volume fractions lead to a much smaller average number of particles entrained per trial, which may be undesirable to create a rapid filtering process. As such, the volume fraction chosen should provide an acceptable balance between filtering efficiency and process duration to meet their individual requirements. 

To provide an efficient sorting by length using a cylindrical substrate, the best combination is to have one fiber population satisfying $L^{*} \leq 1$ and the other one  $L^{*} \geq 1$. This threshold can be tuned by changing the radius $R_{\rm s}$ of the cylindrical substrate. The condition $L^{*} \geq 1$ enforces a preferential alignment that will delay the entrainment of the fiber. Our results also show that, when sorting by length, it is more efficient if the fiber diameter standard deviation is minimized. Given that diameter sorting effects can be of equal or higher order to length sorting, having a wide distribution of diameters allows some undesired fibers to be entrained. 

In this paper, we have outlined a systematic approach that can be used to achieve high-efficiency capillary diameter and length filtering. We should emphasize that various fundamental questions remain open that would allow to increase further the efficiency of the capillary sorting method presented here. For instance, the presence of surfactants is known to modify the streamlines in the liquid bath \cite{Mayer2012}, which could tentatively lead to new sorting effects. Additionally, most capillary sorting and filtering studies have been done with Newtonian fluids, and the extension to non-Newtonian rheology would be interesting to consider. Nevertheless, in summary, the method presented here provides a promising novel soft filtration strategy, which, if designed correctly, has the potential to achieve an efficient selective length or diameter sorting of elongated particles, illustrated here with fibers.

\begin{acknowledgments}
This material is based upon work supported by the National Science Foundation under NSF Faculty Early Career Development (CAREER) Program Award CBET No. 1944844.
\end{acknowledgments}

\bibliography{Bibliography}

\end{document}